\documentclass[reqno,a4paper,11pt]{amsart}

\usepackage{amsmath,amstext,amsfonts,amsbsy,eucal,amssymb}
\usepackage[latin1]{inputenc}

\numberwithin{equation}{section}

\theoremstyle{definition}

\begin{document}

\parskip 4pt
\baselineskip 16pt


\title[On  differential-difference equations admitting Lax representation]
{On  some class of differential-difference equations admitting Lax representation}

\author[Andrei K. Svinin]{Andrei K. Svinin}
\address{Andrei K. Svinin,
Institute for System Dynamics and Control Theory, Siberian Branch of Russian Academy of Sciences,
P.O. Box 292, 664033 Irkutsk, Russia}
\email{svinin@icc.ru}
%
%

\date{\today}



\begin{abstract}
This note is designed to show some classes of differential-difference equations admitting Lax representation which  generalize  evolutionary equations known in the literature.
\end{abstract}

\maketitle

\section{Introduction}

 The simplest case of an evolutionary equation sharing the property of having Lax pair representation  is given by the Volterra lattice
\begin{equation}
r_i^{\prime}=r_i(r_{i-1}-r_{i+1})
\label{Vl}
\end{equation}
which is related via the substitution $r_i=u_iu_{i+1}$ to its modified version
\begin{equation}
u_i^{\prime}=u_i^2(u_{i-1}-u_{i+1}).
\label{mVl}
\end{equation}
In turn, the substitution
\[
r_i=\frac{1}{(v_{i}-v_{i-2})(v_{i+1}-v_{i-1})}
\]
gives the relationship of the Volterra lattice (\ref{Vl}) with one of an equation of Volterra type \cite{Yamilov}
\begin{equation}
v_i^{\prime}=\frac{1}{v_{i+1}-v_{i-1}}.
\label{yVl}
\end{equation}
Remark that all the equations (\ref{Vl}), (\ref{mVl}) and (\ref{yVl}) are evolutionary ones. Moreover these equations admit corresponding hierarchies of generalized symmetries which can be presented in explicit form via some discrete polynomials \cite{Svinin1}, \cite{Svinin2}. It is common of knowledge that  equations (\ref{Vl}), (\ref{mVl}) and (\ref{yVl}) have corresponding integrable generalizations. For example, the equation
\[
r_i^{\prime}=r_i\left(\sum_{j=1}^nr_{i-j}-\sum_{j=1}^nr_{i+j}\right),
\]
known as Itoh-Narita-Bogoyavlenskii lattice \cite{Itoh}, \cite{Narita}, \cite{Bogoyavlenskii}, naturally generalizes the Volterra lattice (\ref{Vl}), while its modified version looks as \cite{Bogoyavlenskii}
\[
u_i^{\prime}
=u_i^2\left(\prod_{j=1}^nu_{i-j}-\prod_{j=1}^nu_{i+j}\right).
\]
Corresponding integrable generalization of equation (\ref{yVl}), namely,
\[
\displaystyle
v_{i}^{\prime}=\frac{1}{\prod_{j=1}^n\left(v_{i-j+n+1}-v_{i-j}\right)}
\]
and explicit form of its hierarchy was given in \cite{Svinin2}. Our main goal of this note is to present further generalization of these equations.

In section \ref{sec:2}, we consider some class of auxiliary linear equations and look for compatibility conditions for these ones. In a result, we get differential-difference equations in a sense admitting Lax pair representation. It is worth remarking that resulting equations, generally speaking, are not of evolutionary type. In section \ref{sec:3}, we consider Darboux transformation of auxiliary linear equations and derive some class of quadratic discrete equations as a condition of compatibility of corresponding linear discrete equations. We observe that obtained equations in a sense generalize known in the literature the lattice potential KdV equation.

\section{Linear equations and its consistency conditions}

\label{sec:2}

\subsection{The first class of differential-difference equations}

Let us consider the following pair of linear equations:
\begin{equation}
zs_i\phi_{i+n}+\phi_i=z\phi_{i+p},\;\;\;
\phi_i^{\prime}=z\xi_i\phi_{i+n},
\label{linear1}
\end{equation}
on some wave function $\phi=\phi_i=\phi_i(x, z)$. By assumption, they are parameterized by some integers $p>n\geq 1$. As is seen, these equations constitute compatible pair provided that two relations, namely,
\begin{equation}
s_i\xi_{i+n}=s_{i+n}\xi_{i+p},\;\;\;
s_i^{\prime}=\xi_{i+p}-\xi_i
\label{pair}
\end{equation}
are valid. Remark that the first relation in (\ref{pair}) can be equivalently rewritten as
\begin{equation}
\prod_{j=1}^{p-n}\xi_{i+j-1}\prod_{j=1}^ns_{i-j}=\delta
\label{1}
\end{equation}
with some arbitrary constant $\delta$. By suitable reparameterization, we can make $\delta =1$. It is a simple observation that putting
\begin{equation}
\xi_i=\prod_{j=1}^nu_{i+j+n-1},\;\;\;
s_i=\prod_{j=1}^{p-n}\frac{1}{u_{i+j+2n-1}},
\label{xiu}
\end{equation}
with some field $u=u_i$, we solve (\ref{1}) with $\delta=1$.  This ansatz, after substituting it in the second relation in (\ref{pair}), gives differential-difference equation
\begin{equation}
\left(\prod_{j=1}^{p-n}\frac{1}{u_{i+j-1}}\right)^{\prime}=\prod_{j=1}^nu_{i-j+p}-\prod_{j=1}^nu_{i-j}
\label{u1}
\end{equation}
which we can rewrite as
\begin{equation}
\sum_{j=1}^{p-n}\frac{u_{i+j-1}^{\prime}}{u_{i+j-1}}=\prod_{j=1}^pu_{i+j-n-1}-\prod_{j=1}^pu_{i+j-1}
\label{u2}
\end{equation}
or in the form
\begin{eqnarray}
\left(\prod_{j=1}^{p-n}u_{i+j-1}\right)^{\prime}&=&\prod_{j=1}^{p-n}u_{i+j-1}\left(\prod_{j=1}^pu_{i+j-n-1}-\prod_{j=1}^pu_{i+j-1}\right) \nonumber \\
                                                &=&\prod_{j=1}^{p-n}u_{i+j-1}^2\left(\prod_{j=1}^nu_{i-j}-\prod_{j=1}^nu_{i+j+p-n-1}\right).
\label{u}
\end{eqnarray}
Relations (\ref{u1}) and (\ref{u2}) can be considered as local differential-difference conservation laws for equation (\ref{u}).

\subsection{The second class of differential-difference equations}

Let us introduce the potential $v_i$ by $s_i=v_{i+p}-v_i$. The relationship between two fields $u$ and $v$, due to (\ref{xiu}), is given by relations
\[
v_{i+p}-v_i=\prod_{j=1}^{p-n}\frac{1}{u_{i+j+2n-1}},\;\;\;
v_i^{\prime}=\xi_i=\prod_{j=1}^nu_{i+j+n-1}.
\]
Therefore  the first relation in (\ref{pair}) becomes
\begin{equation}
\left(v_{i+p-n}-v_{i-n}\right)v_i^{\prime}=\left(v_{i+p}-v_i\right)v_{i+p-n}^{\prime}.
\label{111}
\end{equation}
The latter in fact is equivalent to the differential-difference equation
\begin{equation}
\prod_{j=1}^{p-n}v_{i+j-1}^{\prime}\cdot\prod_{j=1}^n\left(v_{i-j+p}-v_{i-j}\right)=1
\label{3}
\end{equation}

Thus, we have in hand two classes of nonlinear differential-difference equations, namely, (\ref{u}) and (\ref{3}) which gives the compatibility of the linear equations (\ref{linear1}). These equations involve two integers $p>n$ and $n\geq 1$, but one can see that to separate really different equations, one must suppose that $p$ and $n$ are co-prime positive integers. For example, this is the case $p=n+1$, for $n\geq 1$.

\subsection{The third class of differential-difference equations}

Let
\begin{eqnarray}
r_i&\equiv&\frac{\xi_{i-n}}{s_{i-n}}=\frac{\xi_{i+p-2n}}{s_{i-2n}} \label{ri} \\
&=&\frac{v_{i-n}^{\prime}}{s_{i-n}}=\frac{v_{i+p-2n}^{\prime}}{s_{i-2n}} \label{ri1} \\
&=&\prod_{j=1}^pu_{i+j-1}. \label{ri2}
\end{eqnarray}
In virtue of (\ref{3}) and (\ref{ri1}),
\[
\prod_{j=1}^{p-n}r_{i+j-1}=\prod_{j=1}^p\frac{1}{s_{i+j-2n-1}}=\prod_{j=1}^p\frac{1}{(v_{i+j+p-2n-1}-v_{i+j-2n-1})}.
\]
Then
\begin{eqnarray}
\left(\prod_{j=1}^{p-n}r_{i+j-1}\right)^{\prime}&=&-\prod_{j=1}^p\frac{1}{s_{i+j-2n-1}}\cdot\sum_{j=1}^p\frac{s_{i+j-2n-1}^{\prime}}{s_{i+j-2n-1}} \nonumber \\
&=&\prod_{j=1}^{p-n}r_{i+j-1}\left(\sum_{j=1}^p\frac{v_{i+j-2n-1}^{\prime}}{s_{i+j-2n-1}} -\sum_{j=1}^p\frac{v_{i+j+p-2n-1}^{\prime}}{s_{i+j-2n-1}}\right). \nonumber
\end{eqnarray}
Using (\ref{ri1}) we get
\begin{eqnarray}
\left(\prod_{j=1}^{p-n}r_{i+j-1}\right)^{\prime}&=&\prod_{j=1}^{p-n}r_{i+j-1}\left(\sum_{j=1}^pr_{i+j-n-1}-\sum_{j=1}^pr_{i+j-1} \right) \nonumber \\
&=&\prod_{j=1}^{p-n}r_{i+j-1}\left(\sum_{j=1}^nr_{i-j}-\sum_{j=1}^nr_{i+j+p-n-1} \right). \label{INB}
\end{eqnarray}
One can see that (\ref{ri2}) relates two equations (\ref{INB}) and (\ref{u}).

\subsection{Linear equations on new wave function}

Let us introduce $\gamma_i$ such that $\gamma_{i+1}=\gamma_i/u_{i+n}$, then
\[
\gamma_{i+p}=\gamma_{i}\cdot\prod_{j=1}^p\frac{1}{u_{i+j+n-1}}=\frac{s_i}{\xi_i}\gamma_{i}=\frac{\gamma_{i}}{r_{i+n}},
\]
\begin{equation}
\gamma_{i+p-n}=\gamma_{i}\cdot\prod_{j=1}^{p-n}\frac{1}{u_{i+j+n-1}}=s_{i-n}\gamma_{i} \label{sin}
\end{equation}
and
\[
\gamma_{i+n}=\gamma_{i}\cdot\prod_{j=1}^n\frac{1}{u_{i+j+n-1}}=\frac{\gamma_{i}}{\xi_i}.
\]
Let  $\phi_i\equiv \gamma_i\psi_i$. The linear equations  (\ref{linear1}) in terms of new wave function $\psi=\psi_i$ become
\begin{equation}
z\psi_{i+n}+r_{i+n}\psi_i=z\psi_{i+p},\;\;\;
\psi_i^{\prime}=z\psi_{i+n}-\frac{\gamma_i^{\prime}}{\gamma_i}\psi_i.
\label{linear2}
\end{equation}
With (\ref{sin}) and the second equation in (\ref{pair}), we have the following:
\[
\left(\frac{\gamma_{i+p-n}}{\gamma_i}\right)^{\prime}=\frac{\gamma_{i+p-n}}{\gamma_i}\left(\frac{\gamma_{i+p-n}^{\prime}}{\gamma_{i+p-n}
}-\frac{\gamma_i^{\prime}}{\gamma_i}\right)=s_{i-n}^{\prime}=\xi_{i+p-n}-\xi_{i-n}
\]
and then taking into account (\ref{ri}) we get
\begin{eqnarray}
\frac{\gamma_{i+p-n}^{\prime}}{\gamma_{i+p-n}}-\frac{\gamma_i^{\prime}}{\gamma_i}&=&\frac{\xi_{i+p-n}-\xi_{i-n}}{s_{i-n}}=\frac{\xi_{i+p-n}}{s_{i-n}}-\frac{\xi_{i+p-2n}}{s_{i-2n}} \nonumber \\
&=&r_{i+n}-r_i. \nonumber
\end{eqnarray}
We can resolve the latter as
\begin{equation}
\sum_{j=1}^{p-n}\frac{\gamma_{i+j-1}^{\prime}}{\gamma_{i+j-1}}=\sum_{j=1}^nr_{i+j-1}.
\label{4}
\end{equation}
In turn, (\ref{4}) becomes an identity if we substitute
\[
\frac{\gamma_i^{\prime}}{\gamma_i}=\sum_{j=1}^na_{i+j-1},\;\;\;r_i=\sum_{j=1}^{p-n}a_{i+j-1},
\]
with some field $a=a_i$. Then linear equations (\ref{linear2}) become
\begin{equation}
z\psi_{i+n}+a_1^{[p-n]}(i+n)\psi_i=z\psi_{i+p},\;\;\;
\label{linear3}
\end{equation}
and
\begin{equation}
\psi_i^{\prime}=z\psi_{i+n}-a_1^{[n]}(i)\psi_i,
\label{linear4}
\end{equation}
where, by definition, $a_1^{[r]}(i)=\sum_{j=1}^ra_{i+j-1}$, for any integer $r\geq 1$. The consistency conditions for linear equations (\ref{linear3}) and (\ref{linear4}) yield a differential-difference equation
\begin{equation}
\sum_{j=1}^{p-n}a_{i+j-1}^{\prime}=\sum_{j=1}^{p-n}a_{i+j-1}\left(\sum_{j=1}^na_{i-j}-\sum_{j=1}^na_{i+j+p-n-1}\right).
\label{a}
\end{equation}
By direct calculations, one can check that  the ansatz $r_i=\sum_{j=1}^{p-n}a_{i+j-1}$, relates (\ref{a})  to (\ref{INB}).

\section{Darboux transformation}

\label{sec:3}

\subsection{Quadratic discrete equation}

Let us discuss Darboux transformation for linear equations (\ref{linear1}). We consider linear transformation in the form
\begin{equation}
\bar{\phi}_i=\phi_{i+p-n}+g_i\phi_i
\label{10}
\end{equation}
with some coefficient $g_i$ to be defined by condition that (\ref{10}) should be Darboux transformation for (\ref{linear1}). Consider the transformation of the first equation in (\ref{linear1}). We have
\[
z\bar{s}_i\left(\phi_{i+p}+g_{i+n}\phi_{i+n}\right)+\phi_{i+p-n}+g_i\phi_i
=z\left(\phi_{i+2p-n}+g_{i+p}\phi_{i+p}\right)
\]
\[
=zs_{i+p-n}\phi_{i+p}+\phi_{i+p-n}+zg_{i+p}\phi_{i+p}
\]
and therefore
\[
z\left(\bar{s}_i-s_{i+p-n}-g_{i+p}\right)\phi_{i+p}+g_i\phi_i+z\bar{s}_ig_{i+n}\phi_{i+n}=0.
\]
Requiring that (\ref{10}) to be Darboux transformation gives the relations
\[
g_{i+p}-g_i=\bar{s}_i-s_{i+p-n}=\bar{v}_{i+p}-\bar{v}_i+v_{i+p-n}-v_{i+2p-n},\;\;\;
g_is_i=\bar{s}_ig_{i+n}
\]
the first of which is solved by $g_i=\bar{v}_i-v_{i+p-n}$ and therefore the second one is equivalent to the following discrete equation:
\begin{equation}
\left(\bar{v}_i-v_{i+p-n}\right)\left(v_{i+p}-v_i\right)=\left(\bar{v}_{i+p}-\bar{v}_i\right)\left(\bar{v}_{i+n}-v_{i+p}\right).
\label{2}
\end{equation}
Note that this equation in the special case $p=n+1$ appeared in \cite{Svinin2}. One can check, that it can be also written as
\begin{equation}
\left(v_{i+p}-v_i\right)\left(\bar{v}_{i+p}-v_{i+p-n}\right)=\left(\bar{v}_{i+p}-\bar{v}_i\right)\left(\bar{v}_{i+n}-v_i\right)
\label{21}
\end{equation}
and in the form
\begin{equation}
\left(\bar{v}_{i+n}-v_i\right)\left(\bar{v}_i-v_{i+p-n}\right)=\left(\bar{v}_{i+p}-v_{i+p-n}\right)\left(\bar{v}_{i+n}-v_{i+p}\right).
\label{22}
\end{equation}
We observe that replacing $v_i\leftrightarrow \bar{v}_{i}$ and $n\leftrightarrow p-n$ in (\ref{2}), we obtain (\ref{21}). This means that two different pairs of parameters $(p, n)$ and $(p, p-n)$ correspond in fact to the same equation (\ref{2}).

Making use of (\ref{22}), we observe that this quadratic equation has the following integral:
\begin{equation}
I_i=\prod_{j=1}^ng_{i+j-1}\cdot\prod_{j=1}^{p-n}h_{i+j-1},
\label{Ii}
\end{equation}
where $h_i\equiv g_{i+n}+s_i=\bar{v}_{i+n}-v_i$. Remark, that we can also present $I_i$  in the form
\[
I_i=\prod_{j=1}^p\left(\bar{v}_{i+j-1}-v_{i+j+p-n-1}\right).
\]
The latter needs some explanation. This formula involve $\bar{v}_{i+\alpha}$ with $\alpha\in\{0,\ldots, p-1\}$, while $\alpha$ in $v_{i+\alpha}$ is calculated modulo $p$.

\subsection{Some simple examples}

Remark that the equation $I_i=c$ with some constant $c$ in simplest case $n=1$ and $p=2$, namely,
\begin{equation}
\left(\bar{v}_i-v_{i+1}\right)(\bar{v}_{i+1}-v_i)=c
\label{lpkdV}
\end{equation}
is nothing else but lattice potential KdV (lpKdV) equation \cite{Nijhoff1, Nijhoff2} also known as $H_1$ equation in Adler-Bobenko-Suris classification
\cite{Adler}. Therefore one can consider the relation $I_i=c$ with $I_i$ given by (\ref{Ii}) in a sense as a generalization of the lpKdV equation. For example, in the case $p=3$ we have two equations
\[
\left(\bar{v}_i-v_{i+2}\right)(\bar{v}_{i+1}-v_i)(\bar{v}_{i+2}-v_{i+1})=c
\]
for $n=1$ and
\[
\left(\bar{v}_i-v_{i+1}\right)(\bar{v}_{i+1}-v_{i+2})(\bar{v}_{i+2}-v_i)=c
\]
for $n=2$. One sees that these two equations are in fact the same one.

\subsection{Discrete zero-curvature representation for the equation (\ref{2})}

Let $\phi_{k,i}\equiv \phi_{i+k-1}$ for $k=1,\ldots, p$ and $\Phi_i\equiv (\phi_{1,i},\ldots, \phi_{p,i})^T$. Then we can rewrite (\ref{10}) in matrix form $\bar{\Phi}_i=V_i\Phi_i$ or more explicitly as
\[
\bar{\phi}_{1,i}=g_i\phi_{1,i}+\phi_{p-n+1,i},\ldots,
\bar{\phi}_{n,i}=g_{i+n-1}\phi_{n,i}+\phi_{p,i},
\]
\[
\bar{\phi}_{n+1,i}=h_i\phi_{n+1,i}+\frac{1}{z}\phi_{1,i},\ldots,
\bar{\phi}_{p,i}=h_{i+p-n-1}\phi_{p,i}+\frac{1}{z}\phi_{p-n,i}.
\]
Obviously, the second equation which complete discrete zero-curvature representation for (\ref{2}) being of the form $\Phi_{i+1}=U_i\Phi_i$  is explicitly given by the equations
\[
\phi_{k,i+1}=\phi_{k+1,i}\;\;\;\mbox{for}\;\;\;k=1,\ldots, p-1
\]
and
\[
\phi_{p,i+1}=\frac{1}{z}\phi_{1,i}+(v_{i+p}-v_i)\phi_{n+1,i}.
\]
Then, the discrete zero-curvature representation for quadratic equation (\ref{2}) is given by matrix equation $V_{i+1}U_i=\bar{U}_iV_i$.

\subsection{An example. Zero-curvature representation for lpKdV}

Consider simplest case $n=1$ and $p=2$ for which we have following pair of auxiliary linear equations:
\[
\left(
\begin{array}{c}
\phi_{1,i+1} \\
\phi_{2,i+1}
\end{array}
\right)=
\left(
\begin{array}{cc}
0   & 1 \\
\displaystyle
\frac{1}{z} & v_{i+2}-v_i
\end{array}
\right)
\left(
\begin{array}{c}
\phi_{1,i} \\
\phi_{2,i}
\end{array}
\right)
\]
and
\[
\left(
\begin{array}{c}
\bar{\phi}_{1,i} \\
\bar{\phi}_{2,i}
\end{array}
\right)=
\left(
\begin{array}{cc}
\bar{v}_i-v_{i+1}   & 1 \\
\displaystyle
\frac{1}{z} & \bar{v}_{i+1}-v_i
\end{array}
\right)
\left(
\begin{array}{c}
\phi_{1,i} \\
\phi_{2,i}
\end{array}
\right).
\]
Let $\varphi_{1,i}=\phi_{1,i}$ and $\varphi_{2,i}=\left(v_{i+1}-v_i\right)\phi_{1,i}-\phi_{2,i}$. In terms of these new wave functions we have
\begin{equation}
\left(
\begin{array}{c}
\varphi_{1,i+1} \\
\varphi_{2,i+1}
\end{array}
\right)=
\left(
\begin{array}{cc}
v_{i+1}-v_i   & -1 \\
\displaystyle
-(v_{i+1}-v_i)^2-\frac{1}{z} & v_{i+1}-v_i
\end{array}
\right)
\left(
\begin{array}{c}
\varphi_{1,i} \\
\varphi_{2,i}
\end{array}
\right)
\label{zcr}
\end{equation}
and
\[
\left(
\begin{array}{c}
\bar{\varphi}_{1,i} \\
\bar{\varphi}_{2,i}
\end{array}
\right)=
\left(
\begin{array}{cc}
\bar{v}_i-v_i   & -1 \\
\displaystyle
\left(\bar{v}_i-v_i\right)\left(\bar{v}_{i+1}-\bar{v}_i\right)+\left(v_i-\bar{v}_{i+1}\right)\left(v_{i+1}-v_i\right)-\frac{1}{z} & \bar{v}_i-v_i
\end{array}
\right)
\left(
\begin{array}{c}
\varphi_{1,i} \\
\varphi_{2,i}
\end{array}
\right).
\]
If we make use of (\ref{lpkdV}) we obtain
\[
\left(
\begin{array}{c}
\bar{\varphi}_{1,i} \\
\bar{\varphi}_{2,i}
\end{array}
\right)=
\left(
\begin{array}{cc}
\bar{v}_i-v_i   & -1 \\
\displaystyle
-(\bar{v}_i-v_i)^2+c-\frac{1}{z} & \bar{v}_i-v_i
\end{array}
\right)
\left(
\begin{array}{c}
\varphi_{1,i} \\
\varphi_{2,i}
\end{array}
\right).
\]
One can see that the latter is more like (\ref{zcr}). Let $c=\alpha-\beta$ and $1/z=\alpha-\lambda$, where $\lambda$ is a new ``spectral'' parameter. Then, as a result we obtain well-known symmetric zero-curvature representation defined by a pair of equations
\[
\left(
\begin{array}{c}
\varphi_{1,i+1} \\
\varphi_{2,i+1}
\end{array}
\right)=
\left(
\begin{array}{cc}
v_{i+1}-v_i   & -1 \\
\displaystyle
-(v_{i+1}-v_i)^2+\lambda-\alpha & v_{i+1}-v_i
\end{array}
\right)
\left(
\begin{array}{c}
\varphi_{1,i} \\
\varphi_{2,i}
\end{array}
\right)
\]
and
\[
\left(
\begin{array}{c}
\bar{\varphi}_{1,i} \\
\bar{\varphi}_{2,i}
\end{array}
\right)=
\left(
\begin{array}{cc}
\bar{v}_i-v_i   & -1 \\
\displaystyle
-(\bar{v}_i-v_i)^2+\lambda-\beta & \bar{v}_i-v_i
\end{array}
\right)
\left(
\begin{array}{c}
\varphi_{1,i} \\
\varphi_{2,i}
\end{array}
\right).
\]

\section{Conclusion}

In this note we have shown three classes of differential-difference equations (\ref{u}), (\ref{3}) and (\ref{INB}). Equations in these classes are defined by pairs of co-prime integers  $n\geq 1$ and $p>n$. Only in the case $p=n+1$ these equations are evolutionary ones. Here we do not consider the question of constructing of integrable hierarchies associated with auxiliary equation
\[
zs_i\phi_{i+n}+\phi_i=z\phi_{i+p}
\]
or
\begin{equation}
z\psi_{i+n}+r_{i+n}\psi_i=z\psi_{i+p}.
\label{mm}
\end{equation}
We only remember that integrable hierarchies associated with (\ref{mm}) were constructed in explicit form in \cite{Svinin1}. These hierarchies are shown to be directly related to KP flows. Also we constructed explicit form of integrable hierarchy on the field $v=v_i$ in the particular case $p=n+1$ in \cite{Svinin2}. We are going to study this question in more detail in subsequent publications.


\end{document}